# Applying information theory and entropy to eliminate errors in mouse-tracking results

Aleksandar Jevremovic, Panayiotis Zaphiris, Sasa Adamovic, Mati Mottus, Andri Ioannou

**Abstract:**

Mouse-tracking of computer system users represents a less expensive, but also a far more applicable alternative to eye-tracking. The main disadvantage of mouse-tracking are errors manifested as discrepancies between the actual eye-gaze position and the mouse cursor position. This paper presents a method for automated correction of errors arising in mouse-tracking. Our approach draws upon information theory and employs Shannon entropy. The method is based on calculating the entropy of a visual representation of a Web page, i.e., we quantify information potential values of various positions. Information obtained, thereby, is paired with cumulative time intervals, spent by the mouse cursor in each position. In this way, we identify cursor positions that do not match eye-gaze positions. To verify the effectiveness of our method, we compared the eye gaze and mouse cursor heat maps in the following ways: We calculated the coefficient of correlation between the two; we computed Euclidean distance between their centers of gravity; and we performed visual comparison.

# 1 Introduction

Mouse-tracking of computer system users represents a less expensive, but also a far more applicable alternative to eye-tracking. Several researchers have demonstrated high correlation between results obtained with the two methods [15, 28]. One of the main advantages of mouse-tracking is the elimination of expensive and complex equipment used in eye-tracking. Furthermore, doing away with the said complexity has facilitated the use of mouse-tracking in virtually all personal computers and has opened up new avenues for application. Finally, eschewing the equipment associated with eye-tracking and reducing the resulting lab work allows for a more natural behaviour of users participating in a study.

One of its most important contemporary applications of mouse-tracking is in Web environments. Data obtained, thereby, is highly valued as a means of Web content optimisation. Based on this, it is possible to evaluate (determine) various descriptors of Web page content, effectiveness of Web application user interface and the ways users interact with these interfaces. Of course, the target use of collected information is the optimisation of content and user interface. Optimisation can be performed on the basis of user interaction characteristics [3], but also based on the identified user attributes such as gender [4], experience, feelings [5] and negative emotions [6], anxiety [7], ethnic origins [8], bilinguality [9], individual preferences [10], and so forth.

The primary disadvantages of mouse-tracking are errors manifested as discrepancies between the actual eye-gaze position and the mouse cursor position. The magnitude of these errors depends on a multitude of factors: intensity of interaction (e.g., passive activities, such as reading and video watching, which are not mouse-intensive, where correlation with eye-gaze is lower); user profile, and so forth. It is relatively straightforward to identify and manually correct such errors by means of heat maps – one of the most common cumulative representations of data gathered by mouse-tracking and eye-tracking. Despite its simplicity, this process does require significant human time and effort. Additionally, it makes this data unavailable for real-time use (e.g., as an input for artificial intelligence algorithms).

This paper presents a method for automated elimination of errors arising in mouse-tracking. Our approach draws upon information theory and, in particular, employs Shannon entropy. We first calculate the entropy of a visual presentation of a Web page, i.e., we quantify information potential values of various positions. Information obtained, thereby, is paired with cumulative time intervals spent by mouse cursor in each position. In this way, we identify cursor positions that do not match eye-gaze positions. To verify the effectiveness of our method, we compared the eye gaze and mouse cursor heat maps in the following ways: We calculated the correlation coefficient between the two; we computed the Euclidean distance between their centers of gravity; and we performed visual comparison.

## 2 Related work

Using eye-tracking in order to improve the user interface or the human-computer interaction, in general, is a technique that has been studied and used for several decades. This method is not confined to just human-computer interaction and can be applied to any interaction by utilising mobile eye-trackers [13]. Using mouse-tracking as an alternative to eye-tracking is not a novel idea. One of the first studies [14] to examine the correlation between mouse-tracking and eye-tracking was published as early as 2001.

Cooke [15] summarises the main disadvantages associated with eye-tracking as high equipment costs and the possibility of connecting eye movement with certain cognitive processes. This author reports a 69 percent (69%) correlation between eye and mouse movement, thus qualifying mouse-tracking as a usable alternative to eye-tracking. A 2006 paper by Rodden and Fu [16] identifies several patterns of mouse usage during Web search, where position overlapping ranges between six percent (6%) and 70 percent (70%), or 42.2 percent (42.2%) on average.

Franco-Watkins and Johnson [17] introduce a new decision moving-window paradigm that presents stimulus information contingent on eye fixations. They provide data from the first application of this method to risky decision-making and show how it compares to basic eye-tracking and mouse-tracking methods. The authors also enumerate the practical, theoretical, and analytical advantages this method offers above and beyond both mouse-tracking with occlusion and basic eye-tracking of information without occlusion. Finally, they include the use of new metrics that offer more precision than those typically calculated on mouse-tracking data.

Demsar and Çöltekin [18] collected eye and mouse data on tracing of various geometric forms on the screen as a simplification of route tracing on a geographic map. They introduced a new methodology: a combination of space-time visualisation and change detection methods for quantification of eye-mouse interaction in basic tasks.

A 2001 study by Mueller and Lockerd [19] takes into account all mouse movements as a tool for user modeling. They found certain mouse behaviours which are useful for content providers in increasing the effectiveness of their interface design.

In 2001, Ivory and Hearst [20] conducted a survey of usability evaluation methods, organised according to a new taxonomy, that emphasises the role of automation. One of the goals was to identify which aspects of usability evaluation automation are likely to be of use in future research. The paper also suggested new ways to expand existing approaches to better support usability evaluation.

In 2014, Y. Liu et al. [21] conducted a study to collect both the user's feedback on their examinations and the eye-tracking/click-through data while using an experimental search engine. The results revealed that 45.8 percent (45.8%) of the results fixated by users were not recognised by users as being "read", but as being "skimmed". Thus, they proposed a two-stage examination model - "from skimming to reading" stage and a second "from reading to clicking" stage. Additionally, R. W. White and G. Buscher identified text selections as an implicit relevance feedback [22].

In 2012, V. Raghunath et al. tried to evaluate the utility of mouse cursor movement data, in addition to eye-tracking data, in studying pathologists' attention and viewing behaviour. The conclusion of this study was that data detailing mouse cursor movements may be a useful addition to future studies of pathologists' accuracy and efficiency when using digital pathology [23].

In these two studies, V. Navalpakkam et al. identified mouse measures that are strongly correlated with eye movements [1], and developed models to predict user attention (eye gaze) from mouse activity [24, 25].

# 3 Error detection

The fundamental idea for this work was conceived in the process of analyzing mouse movement patterns recorded through the online version of our Computer Networks course textbook. As authors, we were interested in the ways students use (read) the textbook. Therefore, we published the book as a website, with each chapter being a separate webpage. The aim of such approach was to ascertain how interesting certain portions of the content are to students (e.g., passages, images, and formulae). Gaining appropriate insights into user behaviour would enable us to make better choices with regard to which portions of the book should be improved. We used mouse-tracking as the only adequate and widely applicable alternative to eye-tracking. To perform data analysis of recorded mouse positions, we utilised heat maps (Figure 1).

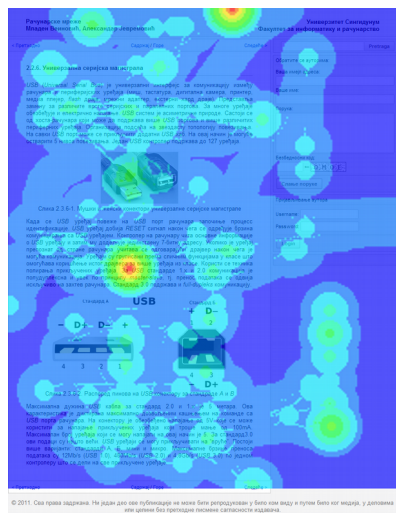

Figure 1. A heat map of a mouse cursor movement for a user reading a chapter in the online textbook

While analyzing the heat maps, we observed the following inconsistency: mouse cursor was spending significantly long intervals in regions with no information value – that is, in regions that contain no useful information for the user (Figure 2). The absence of information in the said regions led us to conclude that the user was unlikely to observe (look at them, gaze at them) them. In other words, we realised that we were most likely dealing with an error manifested as a discrepancy between actual positions of eye gaze and mouse cursor. Therefore, it is natural to assume that users moved the cursor aside, so as to prevent obstruction while reading the text.

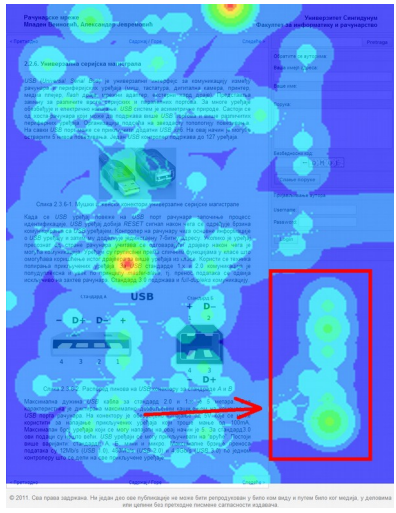

Figure 2. A region without information value where mouse cursor has spent a certain time interval.

Several studies have been published that address the level of correlation between the eye-gaze and mouse cursor positions. Some of these works also provide approximate methods for eliminating the error (see Section: Related work). A downside to some of these approaches is the requirement to classify and group users into certain categories and then make assumptions about eye-gaze positions according to group patterns. An additional disadvantage of this approach (and of the majority of other approaches, as well) is the high complexity of computation and implementation.

The obvious fact that it is unrealistic to assume the user was looking at a completely empty part of the page led us to ask the following question: Is it possible to calculate the information potential of content in certain positions prior to calculating the assumed eye-gaze positions, and then use it to eliminate erroneous (incorrect) eye-gaze positions? It turned out that applying information theory and Shannon entropy, in particular, provided a solution to our question. It is a method with low computational complexity and easy implementation. Our initial trials suggested a high potential for the mentioned and other applications.

## 4 The solution model

Our approach essentially employed the method commonly used to determine the entropy of two-dimensional signals – more precisely Shannon entropy [26]. Entropy is most conveniently defined as an average quantity of information or the measure of uncertainty of an information source. For a known probability, entropy of an event is calculated by:

$$H = -\sum_{i=1}^{l} p_i \log_b p_i$$

where $p_i$ pertains to symbol (pixel's color, or grayscale level, in our case) probabilities obtained through image histograms, and the logarithm base is 2 units.

The value of local entropy varies based on the chosen window size. The window is square-shaped and it represents the number of included neighbouring pixels. The values obtained are represented by means of a binary logarithm, where one bit is the unit of quantity of information. The chosen method allows for the use of a varying number of neighbouring pixels, which is quite similar to Shannon's approximation models – the concept used in modeling natural language as an information source.

The first step in our proposed method was to obtain a two-dimensional matrix with each entry representing an information potential of a pixel, where matrix indices are the (x, y) coordinates of a pixel. Calculated values were normalised in the 0–255 range. Figure 3 provides a visual representation of a matrix in the form of a heat map. Regions with zero information potential (blue color) are easily identified and they directly correspond to page sections without any content. While it is true that the matrix can be further processed by various tools (e.g., clustering), we have opted to stay at the present level for the sake of clarity.

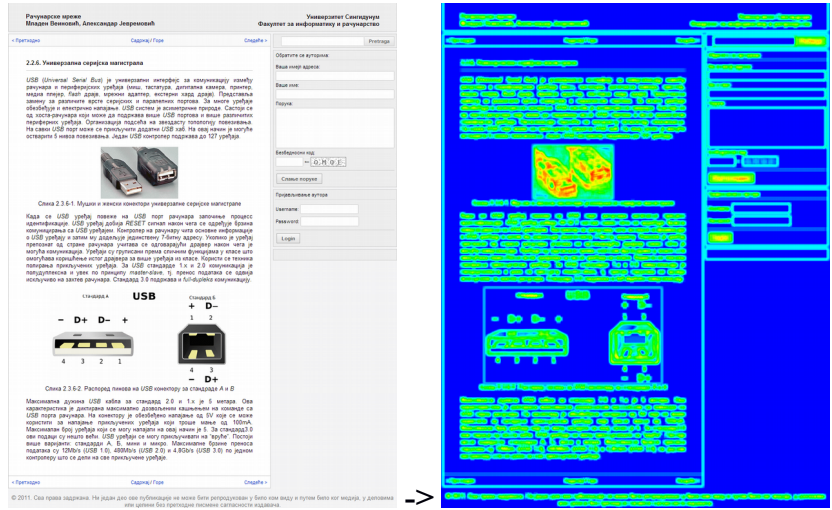

Figure 3. Visual representation of a pixel's information potential.

It should be reiterated that the computational complexity of this method is relatively low, with the time needed to process the enclosed image (1349x1165 pixels) being 1920 milliseconds. We utilised an Intel(R) Core(TM) i5-2450M CPU @ 2.50GHz. We coded in the PHP programming language and used 3x3 pixel squares.

The next step is to pair the matrix of information potentials with the heat map matrix calculated based on the recorded mouse positions. Various methods can be used to this end; but, once again, we elected to use the simplest method – multiplication of corresponding matrix entries, as shown in the formula below:

$$R_{i,j} = H_{i,j} \times I_{i,j}$$

where *R, H,* and *I* are the resultant matrix, heat map matrix, and the matrix of information potentials, respectively.

The main advantage of our choice lies in its simplicity, whereas the disadvantage is in discarding the dwell times spent in positions with zero information potential, which (somewhat) distorts the relative distribution of dwell times. Multiplication results are visually presented in Figure 4.

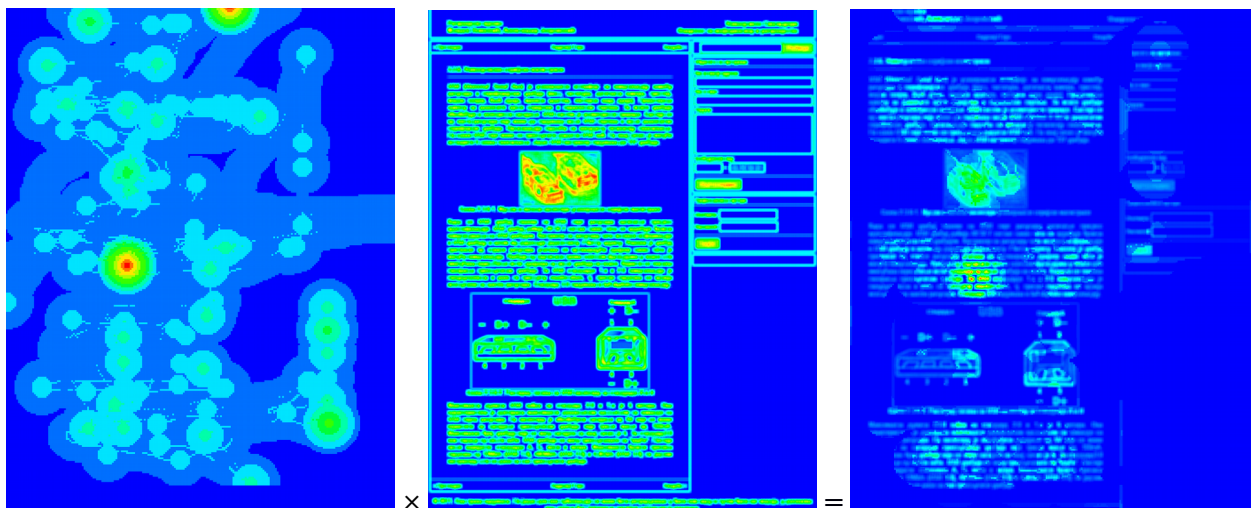

Figure 4. Modification of a heat map obtained by multiplying time sums by information potentials.

Finally, we pair the newly constructed heat map – that is, a map where errors have been removed – with the image of a Web page for which it was constructed (Figure 5). It is clear from Figure 3, that the regions we initially marked as suspicious were automatically removed from the heat map, while values in areas with content (image, text, controls, etc.) were kept.

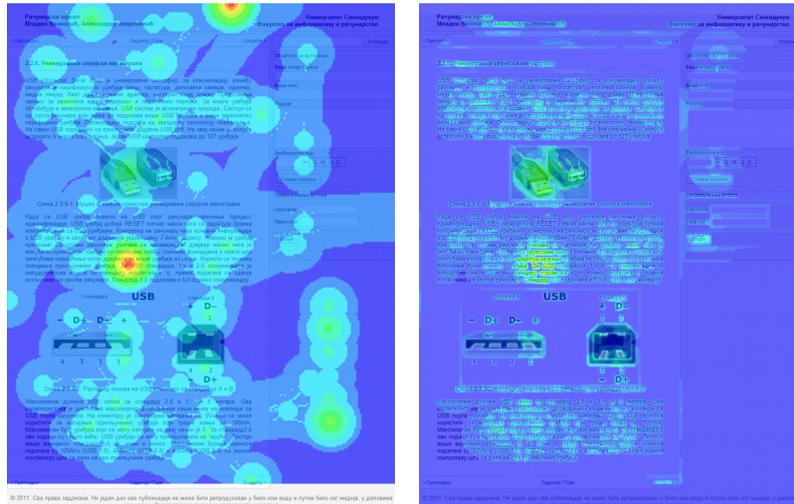

Figure 5. A depiction of a source (left) and a corrected heat map (right).

## 5 Experiment and results

To test the true effectiveness of the proposed method we conducted an experiment of the following characteristics:

- the experiment was carried out in a controlled lab environment;
- 12 individuals took part in the experiment at different time periods;
- total duration of the experiment was 57:23 minutes, or an average of 4:47 minutes per participant;
- all participants used the same Web application on the same desktop computer;
- the Web application consisted of three pages with sections of a detective story and one additional (final) page with pertaining questions (see Fig. 6);
- navigation between pages was implemented with "Prev. page" and "Next page" buttons; and
- over the entire course of the experiment, we simultaneously recorded the eye-gaze positions (with an eye-tracker) and the mouse pointer positions.

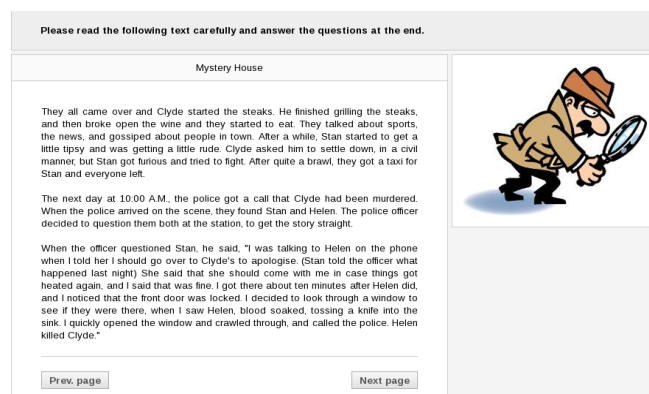

Figure 6. Screenshot of a Web page used in the experiment

We decided to divide the content of the detective story into three pages in order to improve interactivity. Initially, we experimented with the entire story on a single page. This is the case with the minimum expected interaction. However, we observed a behaviour that was not appropriate to the needs of our experiment. Virtually none of the participants used a mouse. In other words, after loading the page, they took their hand off the mouse. For this reason, we started segmenting the story. It turned out that presenting at most two to three paragraphs is the maximum that will keep the users interacting with the mouse. It should be noted that (for this experiment) we purposely had chosen and designed an experiment with an expected lower correlation (text reading) between eye-gaze and mouse cursor positions, which corresponds to a larger error between the two.

## 5.1 Comparison of heat maps

When we speak of a difference in results from direct eye-tracking and mouse-tracking, we also speak of an error. These results are most often used in the form of time series, or summarised, in the form of two-dimensional matrices that are graphically represented as heat maps. Taking into account that heat maps are more prevalent in practice, we decided to measure the effectiveness of the proposed method as similarity (or perhaps dissimilarity) between heat maps obtained by eye-tracking and mouse-tracking (both with and without the error correction method).

There exist different methods and metrics that compare the similarity of values of 2-dimensional matrices. The default metrics, used for this purpose, is the correlation coefficient, which we used in the next chapter (5.2). Additionally, the effectiveness of the proposed method is verified by calculating the distances between the centers of gravity (Chapter 5.3). This metric has not been used in the HCI field until now, but we consider it to have great potential. Finally, the visual representation of the results (obtained by using the different methods) is given in Chapter 5.4.

## 5.2 Correlation coefficient

The correlation coefficient between eye-tracking heat map and mouse-tracking heat map (original, without error elimination) is 0.5352. This correlation is visually represented in Figure 7 (left).

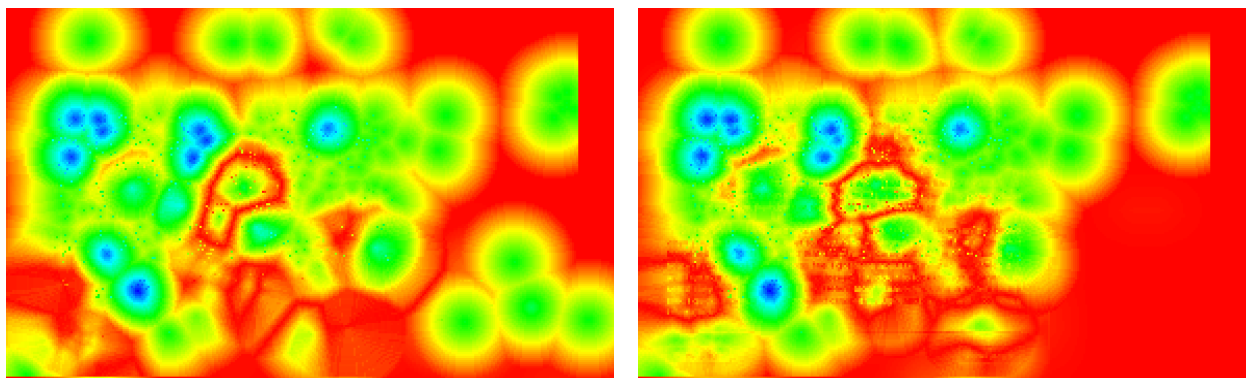

Figure 7. Visual representation of correlations between heat maps

However, after applying the method proposed in this paper, the correlation coefficient is 0.5998. This correlation is visually represented in Figure 7 (right).

## 5.3 Distances between centers of gravity

For the purpose of this work, we made an analogy between a distribution of time (corresponding to attention) and a distribution of mass in a 2D plane (Figure 8). In keeping with this, we chose the center of gravity as a summarised metric of a heat map. As a measure of dissimilarity between two heat maps, we used the Euclidean distance between their centers of mass. The center of gravity is a geometric property of an object and represents an average location of its weight.

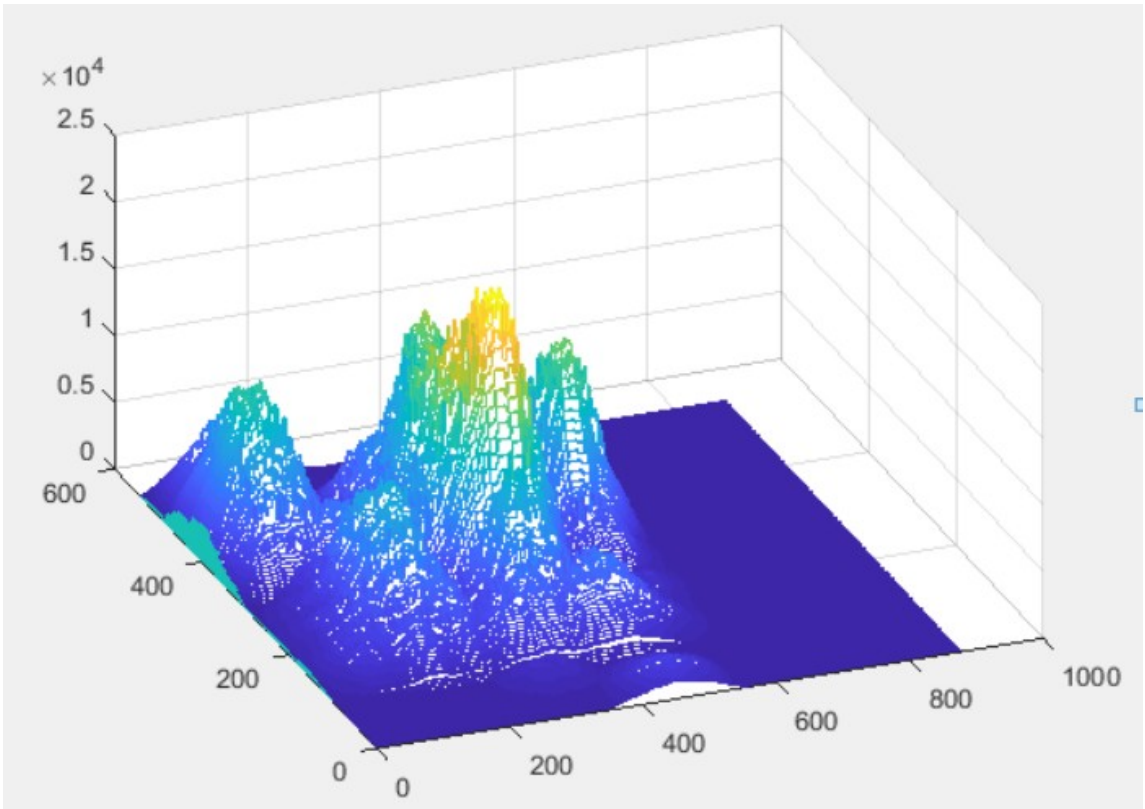

Figure 8. 3D representation of a heat map

The center of gravity is calculated for every dimension of an object in question – that is, for all dimensions of a surface onto which the object is projected. In our case, the center of gravity is calculated for a 2D projection. Therefore, we calculate a horizontal center of gravity:

$$\bar{x} = \frac{m_1 x_1 + m_2 x_2 + .. + m_n x_n}{m_1 + m_2 + .. + m_n}$$

or

$$\frac{\sum_{x=1}^{n_r} m_x \cdot x}{\sum_{x=1}^{r} m_x}$$

and a vertical center of gravity:

$$\bar{y} = \frac{m_1 y_1 + m_2 y_2 + .. + m_n y_n}{m_1 + m_2 + .. + m_n}$$

or

$$\frac{\sum_{y=1}^{n_r} m_y \cdot y}{\sum_{y=1}^{r} m_y}$$

Applying the above formulae, we calculated centers of gravity for various matrices. Table XYZ summarises the results:

Table 1. Centers of gravity coordinates

|    | x   | y   |
|----|-----|-----|
| MT | 406 | 351 |
| ET | 323 | 264 |
| EN  | 376 | 254 |
| EN×MT | 302 | 316 |
| EN×ET | 320 | 263 |

Each of the following abbreviations pertains to a corresponding matrix:

- MT - mouse-tracking heat map
- ET - eye-tracking heat map
- EN - entropy
- EN×MT - entropy applied to mouse-tracking heat map
- EN×ET - entropy applied to eye-tracking heat map

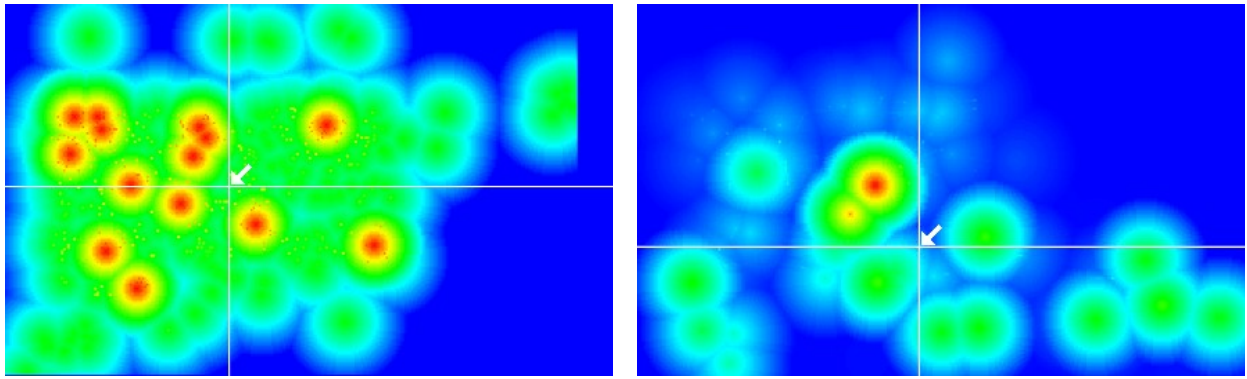

Figure 9. Center of gravity in an eye-tracking heat map (left) and mouse-tracking heat map (right)

Figure 9 depicts a heat map obtained by mouse-tracking, with the center of gravity visibly marked (pixel 406,351).

### 5.3.1 Analysis of effectiveness based on distance between centers of gravity

To assess the similarity between heat maps, we used the Euclidean distance between their centers of gravity as a metric. Euclidian distance represents a distance between two points in Euclidian space [27]. Since we calculate distance in a 2D space, the formula takes the following form:

$$d(p, q) = \sqrt{(p_x - q_x)^2 + (p_y - q_y)^2}$$

where *p* and *q* are the centers of gravity being compared.

The table below provides distances between various centers of gravity:

Table 2. Euclidean distances between different centers of gravity

|       | MT    | ET    | EN   | EN×MT | EN×ET |
| :---: | :---: | :---: | :---: | :---: | :---: |
| MT    | 0     | 120.2 | 101.5 | 109.7 | 123.0 |
| ET    | **120.2** | 0 | 53.9 | 56.1 | 3.2 |
| EN    | 101.5 | 53.9 | 0 | 96.5 | 56.7 |
| EN×MT | 109.7 | **56.1** | 96.5 | 0 | 56.0 |
| EN×ET | 123.0 | 3.2 | 56.7 | 56.0 | 0 |

In order to confirm the initial question of this study, it is crucial to focus on the distance between the eye-tracking center of gravity and mouse-cursor centers of gravity before and after the application of the proposed method. Table 1 shows that the distance between centers of gravity before the method was applied (MT and ET) amounts to 120.2 pixels. Post-application distance (EN×MT and ET) is reduced to 56.1, or half as much. Therefore, we conclude that the heat map obtained from mouse-tracking and subsequently treated with our method is quite similar to the eye-tracking heat map. This means that our proposed method for error reduction in mouse-tracking can be considered effective, at least for the observed task type: reading a text.

The next distance of particular interest is the one between the eye-tracking center of gravity prior to multiplication with the entropy matrix (ET) and after (EN×ET). This value is quite small – only 3.2 pixels. From this, we conclude that the distributions of information potential for eye-gaze positions and dwell times are quite similar. In other words, the distance between gravity centers reveals that the entropy of the visual representation of content can be considered as valid information. For the time being, we limit this conclusion to cases with simple, centralised organisation of content, while we intend to expand our treatment in the future.

## 5.4 Visual comparison of results

Although we envision a chiefly automated application of the proposed method, we also provide a visual depiction of our results for the purpose of verification. Figure 10 shows the original heat map obtained by mouse-tracking (left) and the one after the application of the proposed method (right).

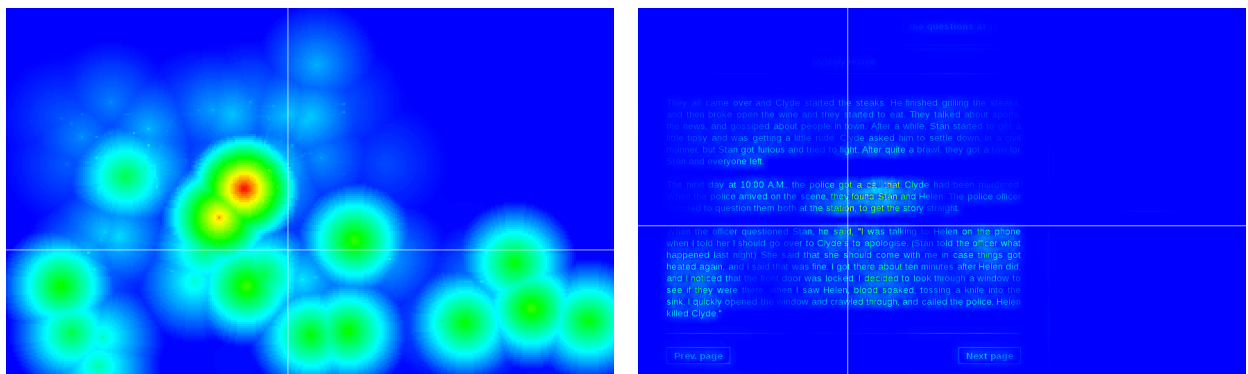

Figure 10. Mouse cursor heat maps before and after applying the proposed method.

In order to make it more human-readable, we applied a blur filter to the corrected heat map (Figure 11).

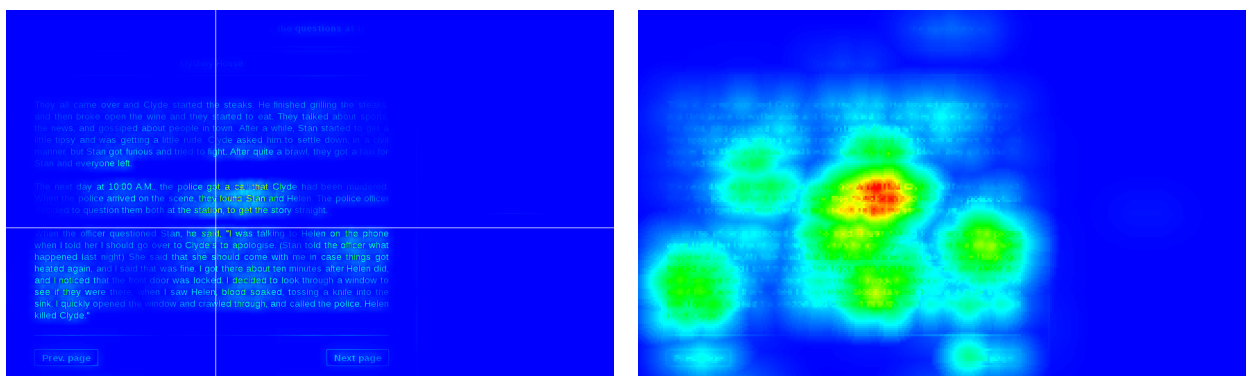

Fig 11. Mouse-tracking heat map, after error correction and blur filter

Figure 12 displays the final results – mouse cursor heat map without error correction (left) and the same map after the error correction method is applied (right).The illustration makes it obvious that the method does not have a significant effect on the zones with content. Conversely, it completely removes the heat from areas without content.

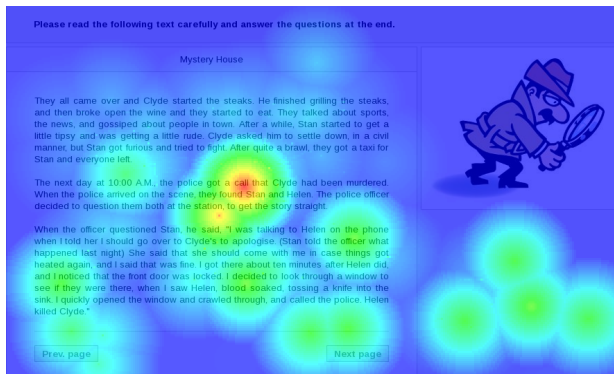
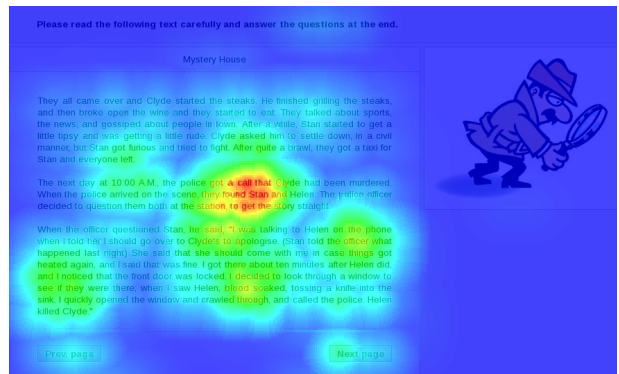

Figure 12. Final result – mouse-tracking heat map before (left) and after (right) applying entropy-based error correction

Next, Figure 13 compares an eye-tracking (left) and a mouse cursor (right) heat map without applying the error correction method. The main difference between these two maps is the lower right region that contains no content. The users have not looked at this area, but have positioned the mouse within.

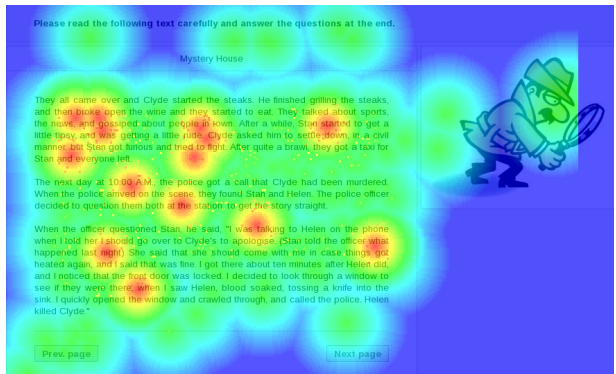
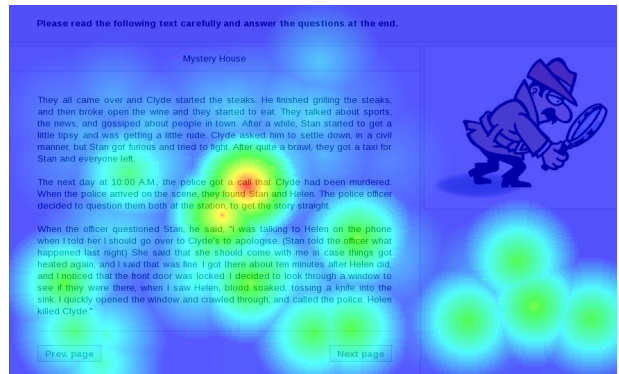

Figure 13. Eye-tracking heat map (left) and mouse-tracking (right) heat map before error correction

Figure 14 puts eye-tracking (left) and mouse-tracking (right) heat maps side by side. Error correction was applied to the latter.

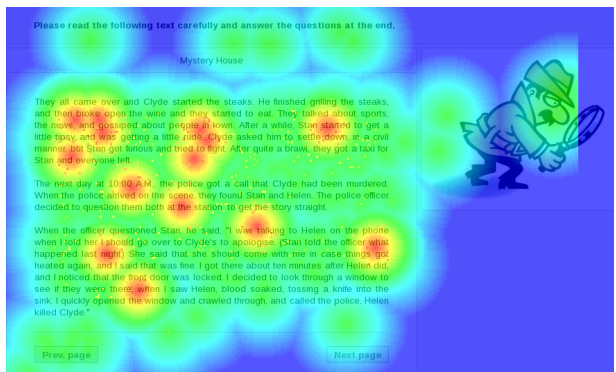
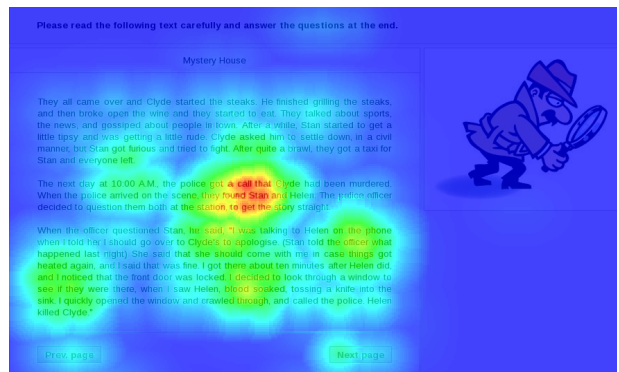

Figure 14. Eye-tracking heat map (left) and mouse-tracking (right) heat map after error correction

Visual comparison of Figures 13 and 14 make for a comfortable conclusion that the (entropy-based) error correction method produces mouse-tracking heat maps with higher similarity to eye-tracking heat maps. In addition, the corrected heat map is more sensible, since it is a more realistic descriptor of a user's behaviour.

# 6   Future work, conclusions

This work presents a method for automated correction of errors occurring in data obtained from mouse-tracking (with the aim of estimating the eye gaze positions. Our method is based on information theory. More specifically, we use the Shannon entropy. Entropy is utilised to determine the information potential of different content regions, i.e. pixels. The matrix of these information potentials we calculate is used to discard the time a mouse cursor spends in positions (pixels) with zero information potential. In this way, we produce a heat map that only contains the time spent in regions that have a non-zero information potential.

We confirmed the effectiveness of the proposed error correction method with an experiment. A comparison of results (heat maps) from simultaneous eye and mouse-tracking was carried out. Firstly, the method application produces a visually-evident, higher correlation. Secondly, this is also confirmed by two quantitative metrics: the correlation coefficient between matrices and the distance between the centers of gravity of compared heat maps.

The use of the center of gravity is an additional contribution of this work. To the extent of our knowledge, we are not aware of any other application of this geometric property in the field of HCI, for measuring correlation coefficient. Our results indicate that there is space to put the property into wider use in this area.

An important result of this study is the entropy-based establishment of connection between the information potential of content and its two-dimensional visual representation. We believe the potential of this link to be much larger than mere elimination of a mouse-tracking error. For example, the analysis of correlation between the distribution of information potentials and the behaviour of a user might serve as one of the inputs into optimisation of both user interface and user experience.

One of the primary directions to improve the proposed method is error correction, instead of its current detection and elimination (discarding). With this goal mind, our method could be supplemented with cluster analysis and the transfer of times spent in regions without information potential to suitable neighbouring regions with information potential. This addition could improve the performance of the proposed solution in scenarios with extremely low or extremely high expected levels of interactivity. Furthermore, accomplishing the said goal would bring the method one step closer to real-time applications, such as time series (analysis).